%% file: main.tex
\documentclass[10pt,conference]{IEEEtran}

\usepackage{xspace}
\usepackage{longtable}
\usepackage{array}
\usepackage[table]{xcolor}
\usepackage{url}
\usepackage{adjustbox}

\usepackage{color}
\usepackage{nameref}
\usepackage{multirow}
\usepackage{tabularx}
\usepackage{censor}
\usepackage{enumitem}
\usepackage[most]{tcolorbox}
\usepackage{mdframed}
\usepackage{fontawesome5}

\definecolor{lightblue}{RGB}{0,0,100}

\newtcolorbox{MyBox}{
  colback=white,
  colframe=lightblue,
  fonttitle=\bfseries,
  coltitle=black,
  sharp corners,
  boxrule=1pt,
  left=5pt,
  right=5pt,
  top=5pt,
  bottom=5pt,
  breakable
}

\newmdenv[
  backgroundcolor=black!6,
  leftline=true,
  rightline=false,
  topline=false,
  bottomline=false,
  linecolor=black,
  linewidth=3pt,
  innerleftmargin=10pt,
  innerrightmargin=10pt,
  innertopmargin=8pt,
  innerbottommargin=8pt,
  skipabove=8pt,
  skipbelow=8pt
]{rqbox}

\newtcolorbox{databox}{
  colback=teal!20!white,
  boxrule=0pt,
  enhanced,
  borderline west={4pt}{0pt}{black},
  left=10pt,
  right=10pt,
  top=8pt,
  bottom=8pt,
  sharp corners
}

\newtcolorbox{resultbox}{
  enhanced,
  colback=blue!5,
  boxrule=0pt,
  sharp corners,
  borderline west={4pt}{0pt}{black},
  left=10pt,
  right=10pt,
  top=8pt,
  bottom=8pt
}

\begin{document}

\title{Understanding LLM Usage Among Early-Career Software Engineers in Practice}

\author{
\IEEEauthorblockN{Julia Alencar}
\IEEEauthorblockA{\textit{CESAR School} \\
Brazil \\
jqa@cesar.school} \\

\IEEEauthorblockN{Cleyton Magalhães}
\IEEEauthorblockA{\textit{UFRPE} \\
Brazil \\
cleyton.vanut@ufrpe.br}

\and

\IEEEauthorblockN{Ronnie de Souza Santos}
\IEEEauthorblockA{\textit{University of Calgary} \\
Calgary, Canada \\
ronnie.desouzasantos@ucalgary.ca}

\and

\IEEEauthorblockN{Italo Santos}
\IEEEauthorblockA{\textit{University of Hawaii at Manoa} \\
United States \\
isantos3@hawaii.edu} \\

\IEEEauthorblockN{Danilo Monteiro Ribeiro}
\IEEEauthorblockA{\textit{CESAR School} \\
Brazil \\
dmr@cesar.school}
}
\maketitle

\begin{abstract}
Despite the rapid adoption of Large Language Models in professional software engineering, limited research has investigated how early career professionals develop effective AI assisted work practices during their transition into industry. We report findings from a mixed methods survey with 75 novice software engineers who actively use LLM supported tools in their daily work. Our results show that LLMs are embedded in routine software engineering activities, including coding, debugging, testing, documentation, and problem solving. Effective use depends on traditional software engineering competencies, such as debugging, testing, and architectural reasoning, together with critical thinking, output verification, prompt engineering, and continuous human oversight. We also identify a gap between workplace expectations and university preparation, with most participants reporting limited formal education on practical LLM use. These findings have implications for software engineering education, organizational onboarding, and workforce development in AI assisted software engineering.
\end{abstract}


\input{intro}
\input{background}
\input{method}
\input{results}
\input{disc}
\input{concl}

\bibliographystyle{IEEEtran}
\bibliography{biblio}

\end{document}

%% file: intro.tex
\section{Introduction}
\label{sec:introduction}
\vspace{-5pt}

Nowadays, Large Language Models (LLMs) are embedded into everyday software engineering workflows through conversational assistants, IDE integrated tools, and AI supported development environments~\cite{ozkaya2023application, hou2024large, vasanthakumar2026impact}. LLM assisted environments support multiple stages of the software development lifecycle, including requirements engineering, design, implementation, testing, deployment, maintenance, and project management~\cite{hou2024large, zheng2025towards, vasanthakumar2026impact}. Developers increasingly rely on these tools for code generation, debugging, documentation, software comprehension, exploratory problem solving, prompt refinement, and validation activities~\cite{rasnayaka2024empirical, tona2024exploring, santos2025model, santana2025software, santos2024we}. Reports from industry indicate that LLM supported tools are now part of many developers' daily workflows~\cite{vasanthakumar2026impact}.

LLM assisted development is associated with faster development, support for repetitive tasks, and perceived productivity gains~\cite{rasnayaka2024empirical, coutinho2024role, vasanthakumar2026impact}. Nevertheless, concerns remain regarding hallucinations, inaccurate outputs, security vulnerabilities, inconsistent responses, and limited contextual awareness~\cite{ozkaya2023application, santos2024we, santana2025software, vasanthakumar2026impact}. Consequently, developers continue to verify, refine, and contextualize AI generated artifacts before incorporating them into software projects, making critical judgment and human oversight central to effective AI assisted software engineering~\cite{santos2025model, santana2025software}.

These changes also have implications for software engineering education and workforce preparation. Existing studies have examined learning with LLMs, overreliance, academic integrity, and students' engagement with AI assisted programming tools~\cite{kirova2024software, tona2024exploring, ferino2025novice, santos2026llm}. However, limited evidence exists on how novice software engineers experience the transition from university to professional environments where LLMs are already embedded in everyday work~\cite{ferino2025novice}. In particular, it remains unclear which technical and non technical competencies early career professionals consider necessary, how they develop effective AI assisted work practices, and how well university education prepares them for workplace expectations. Understanding this transition is important because it represents the point where educational preparation meets professional AI assisted software engineering.

To address this gap, we investigate the following research question: \textbf{RQ. \textit{How are novice software engineers experiencing the integration of LLMs into professional software engineering practice?}} We conducted a mixed methods questionnaire survey with 75 early career software professionals who actively use LLM based tools in their work. Unlike previous studies that focus on educational settings or developers in general, our study specifically investigates novice professionals during their transition into industry. Our findings show that LLMs are already embedded in routine professional activities, but that effective use depends on prompt refinement, output verification, contextual adaptation, critical thinking, and continuous human oversight rather than passive automation. This paper makes four contributions. First, it characterizes how novice software engineers integrate LLMs into professional practice. Second, it identifies the technical and non technical competencies practitioners perceive as necessary for effective AI assisted development. Third, it characterizes the gap between university preparation and workplace expectations. Fourth, it discusses implications for software engineering education, organizational onboarding, and workforce development.

%% file: background.tex
\vspace{-5px}
\section{Related Works: Skills and Practices in AI-Assisted Software Engineering}
\label{sec:background}
\vspace{-5px}

Software engineering requires a combination of technical and interpersonal competencies, including programming, debugging, testing, maintenance, documentation, architectural decision making, communication, teamwork, problem solving, adaptability, and continuous learning~\cite{scaffidi2018employers, borges2024skills}. These competencies support collaboration across technical and non technical stakeholders while enabling software engineers to adapt to evolving technologies and development practices~\cite{scaffidi2018employers, borges2024skills}.

LLM assisted development environments are changing how many of these competencies are exercised. Developers increasingly use conversational systems for code generation, debugging, software comprehension, testing, documentation, and information seeking~\cite{rasnayaka2024empirical, santana2025software, santos2024we, santos2025model}. Development workflows now involve prompt refinement, iterative interaction, and adaptation of generated artifacts to project contexts~\cite{rasnayaka2024empirical, santos2025model}. Although these tools are associated with productivity gains, generated outputs still require review, validation, debugging, architectural assessment, and contextual adaptation before integration into software systems~\cite{ozkaya2023application, santos2025model, vasanthakumar2026impact}. Consequently, software engineers increasingly supervise, evaluate, and refine generated artifacts rather than producing every implementation directly~\cite{ozkaya2023application, vasanthakumar2026impact}.

These changes also introduce competencies related to supervising AI generated work. Developers must construct effective prompts, evaluate generated alternatives, determine when outputs can be trusted, and integrate AI generated artifacts into broader software systems~\cite{hou2024large, zheng2025towards, chen2024opportunities, ozkaya2023application, vasanthakumar2026impact}. Existing studies continue to report concerns regarding hallucinations, inaccurate outputs, security vulnerabilities, inconsistent behavior, overreliance, and reduced independent problem solving when generated content is accepted without sufficient scrutiny~\cite{santos2024we, chen2024opportunities, tona2024exploring, santana2025software, vasanthakumar2026impact}. As a result, critical thinking, verification, code review, debugging, and architectural reasoning remain essential competencies in AI assisted software engineering~\cite{ozkaya2023application, rasnayaka2024empirical, vasanthakumar2026impact}.

These transformations also affect software engineering education and workforce preparation. Students and early career professionals are frequent users of LLM supported tools~\cite{ferino2025novice, kirova2024software, vasanthakumar2026impact}, while concerns regarding overreliance, reduced critical engagement, academic integrity, and independent problem solving continue to emerge~\cite{tona2024exploring, chen2024opportunities, santos2026llm}. Prior work has examined LLM use in software engineering tasks~\cite{santana2025software, santos2024we, santos2025model}, but less is known about how novice professionals transition from university to AI assisted workplaces. This study addresses that gap by examining early career software engineers' perspectives on the skills, practices, and educational preparation needed to work effectively in LLM supported development environments.

%% file: method.tex
\section{Method}
\label{sec:method}
\vspace{-5px}

We conducted a cross-sectional mixed-methods survey~\cite{pfleeger2001principles, easterbrook2008selecting, ralph2020empirical} to examine how novice software engineers use LLM-assisted tools in professional software engineering work. We focus on current practices, perceived technical and non-technical skills, and how early-career professionals view the role of university education in preparing them for AI-assisted development environments. The survey combines closed-ended questions to capture usage patterns and perceptions with open-ended questions to collect reflective accounts of work practices, learning experiences, skill needs, and educational gaps.

\subsection{Survey Design}

In line with established guidelines for survey-based software engineering research~\cite{pfleeger2001principles, linaker2015guidelines, ralph2020empirical}, we designed an anonymous online questionnaire survey implemented using Qualtrics~\footnote{www.qualtrics.com}. The survey combined closed and open-ended questions organized into thematic sections covering professional use of LLMs, technical and non-technical competencies associated with AI-assisted work, perceptions regarding educational preparation, and demographic and professional background information. The instrument was informed by prior discussions regarding AI-assisted software engineering, software engineering competencies, novice developers, and the integration of LLMs into software development workflows~\cite{hou2024large, zheng2025towards, ozkaya2023application, scaffidi2018employers, borges2024skills, ferino2025novice, tona2024exploring, kirova2024software, vasanthakumar2026impact, chen2024opportunities}.

Participants were asked to describe how they use LLMs during software engineering tasks, the steps they follow when interacting with conversational systems, and the skills they consider necessary for effective usage. The survey also included questions regarding prompt refinement practices, evaluation of generated outputs, adaptation of generated artifacts to project constraints, ethical considerations, and self-directed learning activities supported by LLMs. Additional questions investigated how participants perceived the role of university education in preparing them for AI-assisted software engineering environments, including skills learned at university, perceived educational gaps, and recommendations for software engineering curricula. Closed-ended questions included Likert-scale items on practices and perceptions associated with LLM use at work, as well as multiple-choice questions on educational preparation and learning experiences. Open-ended questions enabled participants to provide reflective accounts regarding professional practices, competencies, challenges, and educational experiences (Table~\ref{tab:surveyquestions}). No personally identifiable information was requested. \\ \\

\input{tables/survey}

\vspace{-25px}
\subsection{Pilot}
\vspace{-5px}

Following initial instrument development, the questionnaire underwent a pilot and validation phase before deployment. Two software engineering education researchers with prior experience investigating LLM use in software engineering education reviewed the survey instrument and provided feedback on question clarity, interpretability, relevance, and alignment with the study's goals. Their feedback informed adjustments to wording, question sequencing, and phrasing prior to deployment. The pilot phase also supported refinement of questions investigating technical competencies, non-technical competencies, workplace practices, and educational preparation associated with LLM-assisted software engineering environments.

\subsection{Recruitment}

Participant recruitment combined platform-based sampling and eligibility screening procedures commonly adopted in empirical software engineering research using online participant recruitment platforms~\cite{ralph2020empirical, baltes2022sampling}. Participants were recruited through the Prolific platform, which has been increasingly adopted in software engineering survey research~\cite{russo2022recruiting, reid2022software}. Because identifying novice software engineers directly through platform attributes is difficult, the inclusion criteria combined self-reported educational and professional characteristics. Eligible participants included individuals who reported being software engineering students while working in software-related roles, as well as professionals with up to 2 years of software industry experience. Given the study's focus, active use of LLM-based tools during professional software engineering activities was also required as an inclusion criterion. 

\subsection{Filtering}

Data collection yielded 110 responses. Because identifying novice software engineers solely through recruitment platform filters is difficult, participant eligibility was established using multiple criteria collected in the survey, including professional role, years of software engineering experience, and self reported career level. Participants were considered novices if they reported up to two years of professional software engineering experience and identified themselves as students working in software related roles, junior professionals, or early career software engineers. During screening, 16 participants who self identified as mid level or senior professionals were excluded because they did not satisfy these eligibility criteria, leaving a dataset focused on novice software engineers. The remaining responses underwent a multi stage filtering process to ensure data quality, following recommendations for online survey based software engineering research~\cite{danilova2021you, alami2024you}. Automatic quality filters and attention checks were applied, including domain specific validation questions such as identifying the correct value of a Boolean variable and selecting the description that best matched a compiler's role. Responses containing blank answers, nonsensical text, or random character sequences were removed, reducing the dataset from 110 to 75 responses. Finally, because the survey relied extensively on reflective open ended responses, the remaining submissions were reviewed using an AI generated text detection procedure to identify possible indications of manipulated or automatically generated responses~\cite{de2025investigation}. This step was intended to reduce the inclusion of fabricated qualitative data. No responses were identified as AI generated during this validation.

\subsection{Data Analysis}

The survey produced both qualitative and quantitative data. Closed-ended questions were analyzed using descriptive statistics to summarize reported practices, perceptions, educational experiences, and patterns of LLM usage among novice software engineers~\cite{george2018descriptive}. Given the size of the dataset, the quantitative analysis focused on frequency distributions and descriptive characterization rather than on inferential statistics. Open-ended responses were analyzed using an iterative qualitative coding and categorization process informed by recommendations for qualitative software engineering research~\cite{cruzes2011recommended}. Initially, all responses were read multiple times to support familiarization with the dataset and to identify recurrent meanings, practices, and experiences reported by participants. During the first stage of analysis, relevant excerpts from participants' responses were identified and condensed into core ideas that represent the central meaning of each quotation while preserving the participant's original intent. These core ideas functioned as condensed meaning units that supported subsequent interpretation and comparison across responses. 

In the second stage, the extracted core ideas were interpreted through inductive coding, producing low level codes that represented more specific concepts, actions, competencies, perceptions, or practices described by participants. Two authors independently conducted the coding process. The coding remained inductive and iterative, allowing codes to emerge progressively from the data rather than being predefined before analysis. Similar or conceptually related low level codes were continuously compared and refined throughout the coding process. The resulting codes and categories were regularly discussed with a third author, and disagreements were resolved through consensus meetings. In the final stage, related low level codes were grouped into broader interpretive categories representing higher level patterns observed across the dataset. These interpretive categories varied according to the analytical focus of each survey question. For example, responses regarding professional activities yielded categories associated with LLM usage practices, while questions regarding competencies yielded categories related to technical and non technical skills. Similarly, responses regarding university preparation and teaching approaches yielded categories associated with educational experiences and expectations related to LLM use in software engineering. Table~\ref{tab:codingexample} presents an example of the qualitative coding and categorization process adopted in this study. The qualitative analysis reported in this paper focuses on interpreting and characterizing recurrent themes identified across participants' responses regarding LLM usage practices, competencies, workplace adaptation, and perceptions of educational preparation for AI assisted software engineering environments. The resulting coding data are available in the Data Availability section.

\input{tables/quali}

\subsection{Ethics}

The study was conducted in accordance with institutional guidelines for research involving human participants and received approval from the authors' institutional ethics review process. Participation was voluntary, informed consent was obtained prior to participation, and respondents were informed about the purpose of the study, data handling procedures, and their right to withdraw at any time. No personally identifiable information was collected, and findings are reported only in aggregate and anonymized form.

\section{Threats to Validity}
\label{sec:threats}

Regarding \textit{construct validity}, this study relies on participants' self reported experiences and perceptions of LLM use, which may differ from their actual practices. To mitigate this threat, the survey instrument was informed by prior literature, piloted with two researchers, and combined closed ended and open ended questions. Screening questions, attention checks, and quality filtering procedures were also applied. Regarding \textit{internal validity}, the study adopts a cross sectional design and therefore does not support causal claims. The qualitative analysis involved two authors independently coding the data, with a third author participating in discussions. Disagreements were resolved through consensus meetings, and the resulting coding data are available in the replication package. Regarding \textit{external validity}, participants were recruited through Prolific and represent a convenience sample of novice software engineers who actively use LLM based tools. Although participants came from multiple countries and professional roles, the findings are not intended to be statistically representative of all software engineers and should be interpreted within the context of early career professionals.

%% file: tables/survey.tex
\begin{table*}[t]
\centering
\small
\caption{Survey Questionnaire}
\label{tab:surveyquestions}

\begin{tabular}{p{0.7cm} p{16.5cm}}
\hline

\multicolumn{2}{l}{\textbf{ROLE AND SOFTWARE ENGINEERING TASKS}}\\
\hline

Q1 &
How often do you use LLM-based tools (e.g., ChatGPT, Copilot) in your work?
\newline
( ) Less than once a week \quad
( ) 1--2 times a week \quad
( ) 3--5 times a week \quad
( ) Daily or almost daily.\\

Q2 &
Describe how you use LLM-based tools (e.g., ChatGPT, Gemini, Copilot) in your work activities. You may include examples of common tasks where you use them.\\

\hline

\multicolumn{2}{l}{\textbf{PRACTICES AND SKILLS WHEN WORKING WITH LLMS}}\\
\hline

Q3 &
Describe the steps you follow when using an LLM to complete a task in your work.\\

Q4 &
Technical skills refer to domain-specific knowledge and abilities related to software engineering tasks, tools, and processes. What technical skills do you rely on to use LLMs effectively in your work?\\

Q5 &
Non-technical skills refer to cognitive, interpersonal, and self-regulation abilities that support effective use of LLMs in professional contexts. What non-technical skills do you rely on to use LLMs effectively in your work?\\

\hline

\multicolumn{2}{l}{\textbf{LLM EDUCATION IN SOFTWARE ENGINEERING}}\\
\hline

Q6 &
What did you learn at the university about using LLMs or related AI tools in software engineering?\\

Q7 &
What should software engineering courses teach about using LLMs?\\

Q8 &
Which teaching formats would be effective for learning how to use LLMs in software engineering at university? (Select all that apply.)
\newline
( ) Hands-on projects \quad
( ) Dedicated courses on LLMs \quad
( ) Industry-based projects
\newline
( ) Case studies \quad
( ) Lab sessions \quad
( ) Guided prompt evaluation exercises
\newline
( ) Integration into existing lectures \quad
( ) Guidelines and reading material
\newline
( ) Ethics workshops \quad
( ) Other.\\

\hline

\multicolumn{2}{l}{\textbf{BACKGROUND INFORMATION}}\\
\hline

Q9 & Current professional role.\\

Q10 & Years of professional experience in software engineering.\\

Q11 & Country of residence.\\

Q12 & Gender identity.\\

Q13 & Do you identify as belonging to an underrepresented group in software engineering?\\

\hline

\end{tabular}

\end{table*}

%% file: tables/quali.tex
\begin{table*}[t]
\centering
\small
\caption{Example of Qualitative Coding and Categorization}
\label{tab:codingexample}

\resizebox{\textwidth}{!}{%
\begin{tabular}{p{7.0cm} p{3.2cm} p{3.2cm} p{3.5cm}}
\hline

\multicolumn{4}{l}{\textbf{CODING AND DEVELOPMENT SUPPORT}}\\
\hline

\textbf{Response Excerpt} &
\textbf{Core Idea} &
\textbf{Low-Level Code} &
\textbf{Category} \\
\hline

I used ChatGPT to do coding and sometimes to research something.
&
``used ChatGPT to do coding''
&
Coding support
&
\\

I use LLMs especially for things like generation of the initial structure (boilerplate) in order to save time. I input existing functions to find efficient ways to structure the logic.
&
``generation of the initial structure (boilerplate)''
&
Boilerplate generation
&
\multirow{2}{=}{Coding and Development Support}
\\

Writing code and fixing bugs and testing.
&
``writing code and fixing bugs and testing''
&
Software development support
&
\\

\hline

\end{tabular}
}
\end{table*}

%% file: results.tex
\vspace{-5px}
\section{Results}
\label{sec:results}
\vspace{-5px}

In this section, we report the survey results, covering participant demographics, LLM usage practices, perceived competencies, educational preparation, and gaps between university training and professional practice. We include illustrative quotes reproduced verbatim from participants’ responses to preserve their original wording and support interpretation.

\subsection{Participant Demographics}

The survey included 75 novice software engineering professionals who actively use LLM based tools in their work. Most participants (86.7\%, n=65) reported between one and two years of professional experience, while 13.3\% (n=10) had less than one year, indicating that the sample primarily represents software engineers in the early stages of their careers. Participants represented multiple countries. The largest groups were from Egypt (13.3\%, n=10), India (10.7\%, n=8), and Poland (10.7\%, n=8), followed by Kenya (6.7\%, n=5). Brazil, Germany, Mexico, the Netherlands, Portugal, and Romania each represented 4.0\% (n=3). Additional participants were from Australia, Argentina, Canada, Colombia, France, Italy, Spain, Sweden, Tanzania, the United Kingdom, the United States, and Vietnam. Most participants identified as men (62.7\%, n=47), followed by women (28.0\%, n=21). In addition, 32.0\% (n=24) identified as belonging to groups underrepresented in software engineering, while 58.7\% (n=44) did not. The most frequently reported professional roles were Full Stack Developer (28.0\%, n=21), Backend Developer (17.3\%, n=13), and Data Scientist or Machine Learning Engineer (14.7\%, n=11), followed by QA Engineer (6.7\%, n=5), Frontend Developer (5.3\%, n=4), DevOps Engineer (4.0\%, n=3), and other software engineering related roles (13.3\%, n=10). Seven participants preferred not to disclose demographic information such as country, gender, underrepresented group status, or professional role.

\subsection{How Novice Software Engineers Use LLMs at Work}

Most novice software engineers reported frequently integrating LLM-based tools into their daily professional activities. A total of 58 participants (77.3\%) reported using LLMs daily during work activities, while 13 (17.3\%) reported using them between two and five times per week, and only 4 (5.3\%) reported using them between one and two times per week. These results indicate that LLM-based tools are part of the routine workflows reported by many early-career software professionals. 

\noindent \textbf{Usage.} Participants described using LLMs across a broad range of software engineering activities. The most recurrent behavior involved \textbf{Coding and Development Support} (n=21). Reported activities included code generation, boilerplate creation, frontend implementation, code completion, implementation assistance, and coding-related support practices. One participant explained: \textit{``I use them to generate boilerplate, refactor code or to explain parts of the codebase I'm not familiar with''} (P040). Another recurrent behaviour involved \textbf{Testing and Quality Assurance Support} (n=12). Participants referred to generating unit tests, creating test cases, producing test data, identifying edge cases, supporting validation activities, and assisting quality assurance workflows. One novice software engineer explained: \textit{``It helps me write tests and some easier parts of the code''} (P059). Participants also frequently reported \textbf{Debugging and Troubleshooting} activities (n=10). These activities included understanding unfamiliar errors, troubleshooting technical problems, interpreting logs, and diagnosing implementation issues. One participant stated: \textit{``i paste the error, and it helps me figure out what went wrong''} (P061).

Another reported behaviour involved \textbf{Problem Solving and Reasoning Support} (n=8). Responses referred to discussing alternatives, validating reasoning, brainstorming solutions, planning implementations, and understanding unfamiliar problems. One participant explained: \textit{``I rather try to understand how to think about the issue I'm having''} (P066).

Participants additionally described using LLMs for \textbf{Documentation and Knowledge Support} (n=7). Reported activities included explaining code, summarizing information, generating reports, simplifying technical concepts, and supporting learning unfamiliar technologies. One early-career developer explained: \textit{``They explain code when I don’t understand something and help me find errors''} (P031).

Other reported behaviors included \textbf{Code Maintenance and Improvement} (n=6), \textbf{Productivity and Workflow Support} (n=6), \textbf{Automation Support} (n=3), and \textbf{Specialized Technical Assistance} (n=2). These activities included refactoring existing systems, accelerating workflows, automating repetitive tasks, generating scripts, and assisting with specialized technical tasks.

Participants frequently described using LLMs across multiple stages of their work activities rather than for isolated or occasional tasks. The findings indicate that novice software engineers are integrating LLM-assisted environments into implementation, testing, debugging, learning, reasoning, maintenance, and productivity-related activities as part of their professional practice. \\

\noindent \textbf{Behaviour.} Novice software engineers described different interaction patterns when working with LLM-based tools during software development activities. The most recurrent behavioral pattern involved \textbf{Controlled and Validated Usage} (n=35). Participants described structured workflows involving detailed prompting, contextualization, output review, testing, validation, and iterative refinement before integration into production code or ongoing work activities. One novice software engineer explained: \textit{``I write a clear and specific prompt for the LLM that explains exactly what I want it to do, including all necessary details and context to get an accurate and useful response.''} (P008).

Another recurrent behavioral pattern involved \textbf{Task-Oriented Assistance Usage} (n=15). Participants described using LLMs pragmatically to complete bounded tasks such as obtaining quick solutions, generating code fragments, producing plans, resolving errors, or supporting implementation activities. One participant explained: \textit{``Ask LLM questions. Get answer. Modify answer to fit the project''} (P002).

Participants also frequently reported \textbf{Interactive and Iterative Usage} (n=14). These behaviors involved conversational exchanges, iterative prompt refinement, follow-up questioning, and repeated interaction cycles until acceptable results were achieved. One participant explained: \textit{``write and optimize the prompt to llm, revise and review the output, give the llm feedback and continue''} (P007).

Another behavioural pattern involved \textbf{Exploratory and Agent-Oriented Usage} (n=8). Responses referred to chained prompting, agent-supported workflows, repository-wide contextualization, and integration across multiple AI tools. One participant explained: \textit{``first i identify the problem that i need help with ,then send it to github co pilot or chatgpt and tell him the problem parameters and what i need to solve and we discuss the best approach to take and why and when i am convinced with the solution I/we implement it''} (P012).

Finally, a smaller group of participants reported \textbf{Supportive and Learning-Oriented Usage} (n=3). These behaviours reflected the use of LLMs primarily as supplementary tools for understanding concepts, clarifying errors, or supporting learning-oriented activities. One participant explained: \textit{``I usually try the task myself, and then if i get an error I paste it into Chatgpt and ask why I got that error''} (P016).

\subsection{Skills Required for LLM Usage in Software Engineering}

\noindent \textbf{Technical (Hard) Skills.} Participants described relying on both traditional software engineering competencies and technical skills associated with AI-assisted development workflows. Reported competencies included programming, debugging, software architecture, testing, infrastructure management, prompt engineering, contextual specification, and evaluation of generated outputs. Participants generally perceived effective LLM usage as dependent on existing software engineering expertise combined with competencies associated with interacting with and supervising AI-generated outputs:

\begin{itemize}

\item \textbf{Prompt Engineering and Context Specification (n=30).} This skill involves writing prompts, structuring requests, specifying context, refining instructions, and guiding model behavior. Responses emphasized communicating technical intent clearly and providing contextual information for the LLM. One participant described relying on \textit{``prompt engineering to get the most i need from that LLM''} (P012), reflecting the importance of structuring and contextualizing requests during AI-assisted activities.

\item \textbf{LLM Output Evaluation and Verification (n=27).} This skill involves validating generated outputs, reviewing generated code, identifying hallucinations, evaluating correctness, and maintaining human oversight over generated artifacts. Participants frequently referred to assessing generated outputs before relying on them. As one participant explained, \textit{``reviewing and refining the generated content and understanding the code i checked in the tool fx''} (P006), illustrating verification and oversight activities associated with generated artifacts.

\item \textbf{Debugging and Code Correction (n=23).} This skill involves identifying errors, troubleshooting generated outputs, debugging code, correcting mistakes, and validating software behavior after integration into systems. Responses emphasized the need to carefully inspect and test generated outputs. P061 referred directly to \textit{``debugging skills to test the output''}, illustrating debugging and correction practices during AI-assisted development.

\item \textbf{Software Architecture and System Design (n=17).} This skill involves software architecture, system organization, integration, codebase structure, microservices, and design patterns. Participants emphasized the need to understand how generated code fits within broader software systems. One participant explained that \textit{``A strong grasp of the programming language and system architecture''} is necessary to identify hallucinated or outdated outputs (P014), reflecting the role of architectural reasoning during evaluation of generated code.

\item \textbf{Documentation and Software Engineering Practices (n=9).} This skill involves documentation practices, coding principles, software engineering standards, and code organization activities used to structure and maintain development processes. Responses referred to maintaining organized development practices when working with LLM-generated artifacts. P023 described relying on \textit{``a great knowledge of the sintax and having a document generated to follow easier''}, illustrating documentation and organizational activities associated with software engineering workflows.

\item \textbf{Infrastructure, DevOps, and Development Tooling (n=9).} This skill involves infrastructure management, networking, APIs, cloud services, CI/CD pipelines, version control systems, and development tooling. Participants referred to the need to understand broader technical ecosystems when using LLMs during development activities. One participant referred to the importance of \textit{``basic understanding of software tools to effectively use large language models in my work''} (P069), illustrating tooling and infrastructure related competencies.

\item \textbf{Testing and Quality Assurance (n=6).} This skill involves software testing, quality assurance methodologies, regression testing, smoke testing, and validation of generated artifacts. Responses referred to validating generated outputs and supporting quality assurance activities during development. P009 described activities such as \textit{``Creating test codes, Debuging code, finding errors and solutions, room for optimization, writing plans and code skeletons''}, illustrating testing and QA related practices associated with LLM usage.

\item \textbf{Security and Reliability Practices (n=5).} This skill involves secure development practices, security auditing, dependency management, vulnerability awareness, and protection of sensitive information when interacting with LLMs. Participants referred to concerns regarding security, reliability, and safe integration of generated outputs into development environments. One participant summarized this competency as \textit{``Security Literacy''} (P075), reflecting awareness of security and reliability considerations during AI-assisted work.

\item \textbf{Data and Model Management (n=2).} This skill involves data preparation and selecting appropriate models for specific tasks. Responses referred to managing data and adapting model selection according to task requirements. P029 referred to \textit{``techniques for cleaning data and choosing the right model to use''}, illustrating data and model management practices associated with AI-assisted activities.

\item \textbf{Perceived Lack of Technical Skills (n=1).} This perception reflects participants who did not identify technical skills as necessary for effective LLM usage. For example, P021 stated: \textit{``I do not think I have any technical skills in using LLMs.''}

\end{itemize}

The findings also indicate that these technical competencies rarely appeared independently in participant responses. Participants frequently described combinations of programming knowledge, debugging, prompt engineering, architectural understanding, testing, and output verification within the same responses. The overlap across competencies suggests that effective LLM use in software engineering is associated with combinations of traditional software engineering expertise and competencies for supervising, validating, and integrating AI-generated outputs throughout development. \\

\noindent \textbf{Non-technical (Soft) Skills.} Participants described relying on a broad range of cognitive, interpersonal, and self-management abilities to use LLMs effectively in professional contexts. Reported competencies involved evaluating generated outputs critically, communicating effectively with LLMs, structuring problems, adapting to changing model behavior, and managing iterative interactions during AI-assisted work activities. Participants generally perceived effective LLM usage not only as a technical activity, but also as a process involving reflection, communication, judgment, and self-management:

\begin{itemize}

\item \textbf{Critical Thinking and Evaluation (n=44).} This skill involves reasoning, reflective thinking, evaluating generated outputs, identifying hallucinations, assessing correctness, interpreting responses critically, and questioning the reliability of generated information. Participants frequently emphasized skepticism and human oversight over generated outputs. One participant explained that they rely on \textit{``critical thinking to double-check that the AI's "simple" version hasn't accidentally changed how the logic work''} (P064), suggesting concerns regarding unintended modifications introduced by generated outputs.

\item \textbf{Communication and Perspective Taking (n=42).} This skill involves articulating prompts clearly, refining questions, communicating requirements, expressing ideas precisely, adapting communication styles, and considering user perspectives when interacting with LLMs. Participants frequently emphasized that effective outputs depend on how clearly requests are communicated to the model. For example, P057 emphasized \textit{``Mainly the ability to describe problem clearly (as well as specifications)''}, reflecting the importance participants attributed to prompt articulation and contextual clarity.

\item \textbf{Problem Solving and Analytical Thinking (n=22).} This skill involves decomposing problems, structuring requests, framing tasks, analyzing situations, refining solutions, and reasoning through complex work activities with LLM support. Responses are described using analytical approaches to guide and refine AI-assisted workflows. As one participant noted, \textit{``I use problem-solving skills to ask good questions and understand answers''} (P031), illustrating how analytical reasoning was associated with managing interactions with LLMs.

\item \textbf{Self Regulation and Management (n=16).} This skill involves patience, persistence, multitasking, self-awareness, time management, iterative refinement, and maintaining intentional control over LLM usage. Participants frequently described repeatedly refining prompts and managing interactions until outputs became satisfactory. P043 described this process directly, stating that \textit{``Patience is important because sometimes I need to prompt again and again until the LLM understands and executes the task correctly''}.

\item \textbf{Learning, Creativity, and Adaptability (n=10).} This skill involves adapting to evolving LLM capabilities, continuous learning, experimenting with new interaction approaches, and using creativity during AI-assisted activities. Responses described adjusting workflows to account for differences across models and staying up to date with technological developments. One participant commented that \textit{``All LLMs have a different variation output so i have to adapt to the responses''} (P023), reflecting adaptation to varying model behaviors and outputs.

\item \textbf{Attention to Detail and Carefulness (n=7).} This skill involves carefully reviewing outputs, identifying subtle mistakes, observing inconsistencies, checking formatting and wording, and maintaining precision during interactions with LLMs. Participants emphasized careful inspection of outputs before integrating generated artifacts into work activities. This concern appears in statements such as \textit{``attention to detail to catch mistakes''} (P047), where participants associated careful review with reducing errors and inconsistencies.

\item \textbf{Perceived Lack of Non-technical Skills (n=1).} This perception reflects one participant who did not identify non-technical skills as relevant or necessary for their LLM usage practices. P017 stated: \textit{``I don't. For any activity that requires thinking or interpretation, I do not use LLM.''}

\end{itemize}

The findings suggest that participants framed effective LLM use as an active, human-centered process rather than a passive reliance on automated outputs. Responses referred to supervising generated content, reformulating requests, interpreting ambiguous responses, and refining interactions according to task needs. Participants also emphasized the importance of maintaining judgment and intentional control during AI-assisted activities, suggesting that effective LLM use is associated not only with obtaining outputs from a tool but also with coordinating, guiding, and evaluating interactions throughout professional work.

\subsection{LLM Education in Software Engineering Programs}

\noindent \textbf{Current Educational Coverage.} Novice software engineers described limited educational preparation regarding LLM usage during their university education. A total of 41 respondents indicated that little or nothing was discussed about LLMs in their courses, characterized the coverage as minimal or highly theoretical, or left the question unanswered. Several individuals additionally explained that LLMs were not yet widely available during their studies. 

Among those who reported some exposure to AI-assisted software engineering topics, educational coverage appeared to be concentrated on introductory discussions of AI systems, responsible use, prompting practices, verification of generated outputs, and selected technical aspects of software development workflows. They also referred to discussions on the critical evaluation of generated outputs, the limitations of LLMs, and the risks associated with overreliance on AI-generated artifacts. Although some respondents described exposure to competencies associated with AI-assisted development, the findings suggest that educational preparation remains limited and inconsistent relative to the extent of LLM integration reported in professional practice. Many novice software engineers described learning practical LLM usage strategies independently during work activities rather than through formal software engineering education. \\

\noindent \textbf{Educational Needs for LLM Usage.} When asked what software engineering programs should teach to better prepare students for AI-assisted development environments, responses emphasized practical competencies associated with integrating LLMs into everyday software engineering activities. Novice professionals referred to prompting practices, evaluation and verification of generated outputs, debugging, testing, security awareness, workflow integration, and responsible usage of AI-assisted systems. They also referred to non-technical competencies associated with effective LLM use, particularly critical evaluation, communication, problem-solving, and analytical reasoning. Furthermore, two additional educational expectations emerged specifically regarding university preparation.

The first involved \textbf{Practical Workflow Integration and Productivity Support}. Individuals emphasized that students should learn how to incorporate LLMs into realistic software engineering workflows and professional development activities. One respondent explained that programs should teach students \textit{``how to practically use them to increase your productivity''} (P003), while another referred to the importance of learning \textit{``how to include them in the SDLC more often without affecting a healthy workflow ''} (P012).

The second expectation involved \textbf{Foundational LLM Literacy and Onboarding}. Responses emphasized the need for introductory preparation on what LLMs are, how they function, and how students should begin interacting with them before more advanced AI-assisted practices are introduced. One novice software engineer explained that courses should \textit{``teach how to use LLMs from scratch''} and \textit{``don't consider the students already have a background''} (P008). Another respondent referred to the importance of teaching \textit{``basic concepts about LLMs and how to use it appropriately''} (P021). \\

\noindent \textbf{Preferred Teaching Formats.}
Respondents were also asked which educational formats they perceived as effective for teaching LLM usage in software engineering programs. The findings suggest a preference for applied, practice-oriented educational approaches that involve direct interaction with AI-assisted development activities, realistic software engineering workflows, and guided evaluation of generated outputs.

\begin{itemize}

\item \textbf{Hands-on projects (n=57).} Activities involving direct usage of LLMs during software development tasks, implementation activities, and practical exercises.

\item \textbf{Dedicated courses on LLMs (n=40).} Standalone university courses specifically focused on LLMs, AI-assisted software engineering, and practical AI usage in development environments.

\item \textbf{Industry-based projects (n=39).} Collaborations with industry, real-world development scenarios, and exposure to professional software engineering practices involving LLM usage.

\item \textbf{Case studies (n=39).} Analysis of practical examples, software engineering scenarios, implementation situations, and real-world cases involving AI-assisted development.

\item \textbf{Lab sessions (n=37).} Supervised practical sessions involving experimentation, guided usage, technical exercises, and structured interaction with LLM-based tools.

\item \textbf{Guided prompt evaluation exercises (n=36).} Activities involving prompt writing, refinement, output comparison, verification, and evaluation of generated responses.

\item \textbf{Integration into existing lectures (n=29).} Incorporating LLM-related discussions and activities into traditional software engineering courses and existing curricular structures.

\item \textbf{Guidelines and reading material (n=17).} Supporting material such as documentation, tutorials, recommendations, instructional resources, and written guidance regarding effective and responsible LLM usage.

\item \textbf{Ethics workshops (n=15).} Dedicated discussions and workshops regarding ethics, responsible AI usage, misuse risks, privacy concerns, and professional implications associated with AI-assisted software engineering.

\end{itemize}

A total of seven respondents did not provide responses to this question.

\subsection{Answer to the Research Question}

Considering both the workplace experiences of novice software engineers and their reported educational preparation, our study helps answer our RQ: \textit{How are novice software engineers experiencing the integration of LLMs into professional software engineering practice?} Our findings suggest that novice software engineers experience the integration of LLMs while simultaneously adapting to their first professional roles. Participants described incorporating LLMs into implementation, debugging, testing, documentation, reasoning, and learning activities while developing practices for prompt refinement, output verification, contextual adaptation, and continuous human oversight. Effective participation was perceived to depend on both traditional software engineering competencies and reflective practices for the critical evaluation and responsible use of AI generated artifacts. Participants also perceived a gap between workplace expectations and university preparation for AI assisted software engineering. Many reported limited exposure to practical LLM use during their studies and described acquiring AI assisted development practices independently during their first professional experiences. They emphasized the importance of practical learning opportunities that develop competencies for supervising, evaluating, and integrating AI generated artifacts into software engineering workflows. Our findings indicate that the transition into AI assisted software engineering involves more than learning new tools. For the novice software engineers in our study, it also involved developing professional practices that combine traditional software engineering competencies with supervision, verification, and contextual judgment. By focusing on the transition from university to professional practice, this study extends current knowledge by providing empirical evidence on how early career software engineers develop AI assisted work practices during their first years in industry.

%% file: disc.tex
\section{Discussion}
\label{sec:discussion}

This section discusses our findings in relation to existing literature and their implications for software engineering research, education, and practice.

\subsection{Comparing Results with the Literature}
\label{sec:discussioncomparing}

Our findings align with existing literature describing the integration of LLM assisted environments into software engineering workflows spanning implementation, debugging, testing, documentation, software comprehension, and problem solving~\cite{hou2024large, zheng2025towards, ozkaya2023application, vasanthakumar2026impact}. Participants described conversational and iterative practices, including prompt refinement, output validation, contextual adaptation, and sustained human oversight, consistent with prior work on AI assisted development workflows~\cite{santos2025model, rasnayaka2024empirical}. Our results also corroborate concerns regarding hallucinations, inaccurate outputs, inconsistent responses, overreliance, and the need to verify generated artifacts before integration into software systems~\cite{santana2025software, santos2024we, chen2024opportunities, tona2024exploring}. Together, these findings reinforce that effective AI assisted software engineering continues to depend on supervision, contextual evaluation, and validation.

Our findings also support previous arguments that AI assisted development reshapes rather than replaces traditional software engineering competencies~\cite{ozkaya2023application, vasanthakumar2026impact}. Participants continued to rely on programming, debugging, testing, architectural reasoning, infrastructure knowledge, and communication while also engaging in prompting, output evaluation, and contextual adaptation. This reinforces prior discussions that analytical reasoning, critical thinking, and continuous learning remain essential because generated outputs still require interpretation and judgment~\cite{chen2024opportunities, rasnayaka2024empirical, tona2024exploring, borges2024skills}. Consistent with previous work, participants also described using LLMs to support learning, software comprehension, and exploration of unfamiliar technologies, while recognizing the risks of reduced critical engagement and overreliance~\cite{santos2025model, rasnayaka2024empirical, ozkaya2023application, chen2024opportunities, tona2024exploring, santos2026llm}.

The main contribution of this study extends beyond confirming existing AI assisted development practices. While prior work has explored productivity, technical limitations, educational concerns, and prompting practices~\cite{hou2024large, zheng2025towards, ozkaya2023application, vasanthakumar2026impact, tona2024exploring, chen2024opportunities, santos2026llm}, comparatively little is known about how novice software engineers develop these practices during the transition from university to professional environments~\cite{ferino2025novice}. Our findings show that participants often learned AI assisted development practices independently during their first professional experiences rather than through formal university instruction. They also emphasized the need for practical preparation through realistic software engineering scenarios, hands on projects, and guided evaluation activities. These findings suggest that AI assisted software engineering is not simply a matter of learning new tools. Instead, successful participation depends on developing reflective practices that combine traditional software engineering competencies with supervision, contextual interpretation, verification, and responsible integration of generated artifacts. This perspective extends existing literature by characterizing the transition into AI assisted professional practice rather than AI assisted development itself.

\subsection{Implications}

\noindent \textbf{Research.} This study expands current knowledge by characterizing how novice software engineers develop AI assisted work practices while transitioning into professional environments, a stage that remains largely unexplored in software engineering research. Our findings suggest several directions for future work. Longitudinal studies can investigate how competencies such as prompt engineering, verification, and contextual judgment evolve with professional experience. Comparative studies can investigate differences between novice and experienced engineers, or between organizations with different levels of AI adoption and governance. Future research may also investigate how organizational policies, mentoring, and team practices influence trust, reliance, and effective collaboration with LLM supported tools. Finally, our findings provide a basis for developing and evaluating interventions that support the transition into AI assisted software engineering, including educational approaches, onboarding strategies, and organizational guidance. \textbf{Education.} Our findings suggest that software engineering education does not yet fully reflect AI assisted professional practice. Many participants reported limited exposure to practical LLM use at university despite relying on these tools in daily work. Preparing students for these environments requires more than technical instruction; it also requires developing competencies in verification, contextual reasoning, and responsible integration of generated artifacts. Participants emphasized hands on projects and realistic development scenarios as effective ways to develop these competencies. \textbf{Practice.} Although participants frequently used LLM supported tools, they also described the need to continuously verify outputs, refine prompts, assess generated artifacts, and maintain contextual awareness. Most reported learning these practices independently after entering industry, suggesting opportunities for organizations to better support early career professionals. Structured onboarding, mentorship, code review practices that explicitly evaluate AI generated artifacts, and organizational guidelines for appropriate LLM use may foster more effective and responsible AI assisted development while reducing overreliance and supporting professional growth.

%% file: concl.tex
\vspace{-5px}
\section{Conclusion}
\label{sec:conclusion}
\vspace{-5px}

We investigated how novice software engineers experience LLM-assisted environments in professional software engineering practice. Our findings show that LLM-supported tools are already part of early-career workflows, supporting coding, debugging, testing, documentation, and learning. Participants described LLM use as an active process involving prompt refinement, output verification, contextual adaptation, and critical judgment, rather than simple automation. Effective use, therefore, depends on both traditional software engineering competencies and emerging skills related to supervising and evaluating generated outputs. At the same time, participants reported a gap between workplace expectations and university preparation, often having to learn to use LLMs independently after entering the industry. These findings contribute to discussions on AI-assisted software engineering and suggest implications for education, onboarding, and workforce development. \textbf{Future Work.} We plan to expand the survey across different roles, backgrounds, and workplace contexts, and to conduct interviews with novice engineers to better understand the challenges reported in this study. These next steps aim to inform educational and industry guidance for preparing early-career professionals for AI-assisted software engineering practice.

\vspace{-10px}
\section{Data Availability}
The data used in this study is available at: \url{https://figshare.com/s/7bce6e9ac5ab2383a9d2}
\vspace{-5px}